\documentclass[useAMS,usenatbib]{mn2e}
\usepackage[english]{babel}

\usepackage{fancyhdr}
\usepackage{amsfonts}
\usepackage{amsmath}
\usepackage{amssymb}
\usepackage{multicol}
\usepackage{layout}
 \usepackage{psfig}

\usepackage{graphicx}
\usepackage{subfigure}

\usepackage{times}
\usepackage{natbib}

\newif\ifAMStwofonts
\AMStwofontstrue

\renewcommand{\vec}[1]{\bmath{#1}}
\newcommand{\be}{\begin{equation}}
\newcommand{\ee}{\end{equation}}
\newcommand{\ba}{\begin{eqnarray}}
\newcommand{\ea}{\end{eqnarray}}
\newcommand{\brr}{\begin{array}}
\newcommand{\err}{\end{array}}
\newcommand{\bc}{\begin{center}}
\newcommand{\ec}{\end{center}}

\newcommand{\mincir}{\raise
  -2.truept\hbox{\rlap{\hbox{$\sim$}}\raise5.truept \hbox{$<$}\ }}
\newcommand{\magcir}{\raise
  -2.truept\hbox{\rlap{\hbox{$\sim$}}\raise5.truept \hbox{$>$}\ }}
\newcommand{\siml}{\raise
  -2.truept\hbox{\rlap{\hbox{$\sim$}}\raise5.truept \hbox{$<$}\ }}
\newcommand{\simg}{\raise
  -2.truept\hbox{\rlap{\hbox{$\sim$}}\raise5.truept \hbox{$>$}\ }}

\newcommand{\aj}{AJ}
\newcommand{\apj}{ApJ}

\newcommand{\apjs}{ApJS}
\newcommand{\aap}{A\&A}

\newcommand{\mnras}{MNRAS}

\newcommand{\physrep}{Phys. Rep.}
\newcommand {\apgt} {\ {\raise-.5ex\hbox{$\buildrel>\over\sim$}}\ }
\newcommand {\aplt} {\ {\raise-.5ex\hbox{$\buildrel<\over\sim$}}\ }

\title[KSB: biases and corrections]
  {Biases in, and corrections to, KSB shear measurements}
\author[Viola et al.]
  {M. Viola, P. Melchior, M. Bartelmann\\
  Zentrum f\"ur Astronomie, ITA, Universit\"at Heidelberg, Albert-Ueberle-Str. 2, 69120 Heidelberg, Germany\\
  (mviola,pmelchior,mbartelmann@ita.uni-heidelberg.de)}

\date{\today}

\begin{document}
\label{firstpage}
\maketitle

\begin{abstract}

We analyse the KSB method to estimate gravitational shear from surface-brightness moments of small and noisy galaxy images. We identify three potentially problematic assumptions. These are: (1) While gravitational shear must be estimated from averaged galaxy images, KSB derives a shear estimate from each individual image and then takes the average. Since the two operations do not commute, KSB gives biased results. (2) KSB implicitly assumes that galaxy ellipticities are small, while weak gravitational lensing assures only that the change in ellipticity due to the shear is small. (3) KSB does not invert the convolution with the point-spread function, but gives an approximate PSF correction which -- even for a circular PSF -- holds only in the limit of circular sources. The effects of assumptions (2) and (3) partially counter-act in a way dependent on the width of the weight function and of the PSF. We quantitatively demonstrate the biases due to all assumptions, extend the KSB approach consistently to third order in the shear and ellipticity and show that this extension lowers the biases substantially. The issue of proper PSF deconvolution will be addressed in a forthcoming paper.

\end{abstract}


\section{Introduction}

Cosmic shear measurements are a potentially powerful probe of structure growth at intermediate to late epochs of the cosmic history. Sufficiently precise measurements should be able to constrain both the amount, distribution and fluctuation amplitude of dark matter and the time evolution of dark energy \citep[see][for a recent review]{Bartelmann10}. Significant cosmic-shear signals have been detected in many studies \citep[see][for recent examples]{Bacon00, Kaiser00,Benjamin07}, and evidence for accelerated expansion \citep{Schrabback09} has also been found. Cosmic-shear measurements are the primary motivation for several dedicated surveys that are proposed or upcoming (e.g. EUCLID\footnote{http://sci.esa.int/euclid}, JDEM\footnote{http://jdem.gsfc.nasa.gov}, DES\footnote{http://www.darkenergysurvey.org}, LSST\footnote{http://www.lsst.org}). They will greatly increase the survey area and the number of observed galaxies, and hence lower the statistical uncertainty of the shear estimates. However, the analysis of synthetic data in the GREAT08 challenge \citep{Bridle09} shows that the accuracy of shear estimation methods is currently not sufficient to fully exploit the next-generation surveys \citep{Amara08}.

Particularly concerning are systematic biases in shear estimates, which do not vanish when averaged over a large ensemble of lensed galaxies. These biases often stem from assumptions made in the derivation or implementation of shear estimation methods, which do not hold in reality. For instance, the models used to describe the galactic shape may systematically differ from the true shape \citep{Lewis09,Voigt10,Melchior10}. Model-independent approaches may therefore be favored as they are not limited by the peculiarities of an underlying model. The widely employed KSB method \citep{KSB95} is model-independent because it expresses the lensing-induced shape change by combinations of moments of the galactic light distribution. However, it relies on several assumptions regarding (1) the strength of the apparent distortion, (2) the width of the window function, (3) the mapping between convolved and unconvolved ellipticity, and (4) the ellipticity of the PSF.

It has been noted that the accuracy of the KSB method has a problematic dependence on items (2) and (4) \citep{KK99,Erben01,Hoekstra98}. In this work we investigate the reasons for problems encountered with shear estimates from KSB by a rederivation of its fundamental relations. We take particular care of inspecting the assumptions made, and show if and how improvements to the original KSB relations can be incorporated such that the shear estimates remain free of bias in a wider range of galactic and PSF parameters.

In Sect.~2, we briefly review the basic relations of gravitational lensing and show how to construct a shear estimate from observed image ellipticities. In Sect.~3, we introduce and test three variants of KSB based on a linearized relation between shear and ellipticity, and a novel one which employs a third-order relation. We investigate the validity of the PSF-correction approach in Sect.~4 and comment on the possibility of an improved correction for PSF ellipticities. We conclude in Sect.~5.

\section{Weak lensing basics}\label{sec:weaklensing}

This section summarises the basic weak-lensing concepts that will be used later. For a complete overview we refer to \cite{Bartelmann01}. An isolated lens with surface mass density $\Sigma(\vec{\theta})$ has the lensing potential
\begin{equation}
\Psi(\vec{\theta})=\frac{4G}{c^2}\frac{D_{l}D_{s}}{D_{ls}}\int d^2\theta^{\prime}\Sigma(\vec{\theta}^{\prime})\ln |\vec{\theta}-\vec{\theta}^{\prime}|, \label{eq:potential}
\end{equation}
where $G$ and $c$ are the usual constants and $D_{l, s, ls}$ are the angular-diameter distances between the observer and the lens, the observer and the source, and the lens and the source, respectively.

To sufficient accuracy, light rays are deflected by the angle
\begin{equation}
\vec{\alpha}(\vec{\theta})=\nabla \Psi(\vec{\theta})\;,\label{eq:deflection}
\end{equation}
which relates the angular positions of the source $\vec\beta$ and the image $\vec\theta$ on the sky by the lens equation
\begin{equation}
\vec{\beta}=\vec{\theta}-\vec{\alpha}(\vec{\theta})\;.\label{eq:lenseq}
\end{equation}
If the lens mapping changes little across the solid angle of a source, the lens mapping can be locally linearised to describe the image distortion the Jacobian matrix
\begin{equation}
A\equiv\frac{\partial\vec{\beta}}{\partial \vec{\theta}}=\left(\delta_{ij}-\frac{\partial ^2\Psi(\vec{\theta})}{\partial \theta_{i}\partial \theta_{j}}\right)=\left(\begin{array}{cc}
1-\kappa -\gamma_1 & -\gamma_2 \\
-\gamma_2 & 1-\kappa + \gamma_1 \end{array} \right)\;, \label{eq:distortion}
\end{equation}
with the convergence
\begin{equation}
\kappa(\vec{\theta})=\frac{1}{2}(\Psi_{11}+\Psi_{22})\label{eq:kappa}
\end{equation}
and the two components
\begin{equation}
\gamma_1=\frac{1}{2}(\Psi_{11}-\Psi_{22})\;,\quad\gamma_2 = \Psi_{12}
\end{equation}\label{eq:shearcomp}
of the complex shear $\gamma=\gamma_1+i\gamma_2$. Image distortions measure the reduced shear
\begin{equation}
g=\frac{\gamma}{1-\kappa}
\end{equation}
instead of the shear $\gamma$ itself. To linear order, $\vec\theta$ and $\vec\beta$ are related by
\begin{equation}
\beta_{i}=A_{ij}\theta^{j}\;.
\label{eq:coordtransf}
\end{equation} 

\subsection{Shear estimation}

The shape of an extended source can be descibed by angular moments of its surface brightness distribution $I(\vec{\theta})$,
\begin{equation}
Q_{ij...k}=\int I(\vec{\theta})\theta_{i}\theta_{j}...\theta_{k}d^2\theta\;.\label{eq:moment}
\end{equation}
$Q$ is the total flux, $Q_i$ defines the centroid of the image, and higher-order moments provide information on the image's morphology. Combinations of second moments are used to quantify the image's ellipticity, which we introduce as
\begin{equation}
\chi=\frac{(Q_{11}-Q_{22})+2iQ_{12}}{Q_{11}+Q_{22}}\;.
\label{eq:ellipticity}
\end{equation}
The complex ellipticity $\chi$ is related to the reduced shear $g$ by
\begin{equation}
\chi^{s}=\frac{\chi - 2g+g^{2}\chi^{*}}{1+|g|^2-2\Re(g\chi^{*})}\;,
\label{eq:chi_g}
\end{equation}
\citep{Schneider95}, where $\chi^s$ is the unlensed (intrinsic) ellipticity. This relation holds as long as the lens mapping can be locally linearised. Information on the intrinsic ellipticity of a single object is not accessible. Reasonable shear estimates thus require averaging over many galaxies in a region where $g$ can be considered constant, assuming that the average of $\chi^{s}$ vanishes,
\begin{equation}
0=\langle \chi^{s} \rangle=\left\langle
\frac{\chi - 2g+g^{2}\chi^{*}}{1+|g|^2-2\Re(g\chi^{*})}
\right \rangle\;.\label{eq:gavetot}
\end{equation}
If the coordinate frame is rotated such that only one shear component does not vanish, Eq.~(\ref{eq:gavetot}) is solved by
\begin{equation}
g \simeq \frac{\langle \chi \rangle}{2(1-\sigma^{2}_{\chi})}+ \frac{\langle \chi \rangle^{3}}{8}\frac{1-5\sigma^{2}_{\chi}}{(1-\sigma^{2}_{\chi})^{4}} + \mathcal{O}(\langle \chi\rangle^5)\label{eq:gave}
\end{equation}
where $\sigma_{\chi}$ is the standard deviation of the intrinsic ellipticity distribution. In the derivation of the equation above we neglected higher order moments of the intrinsic ellipticity distribution.
Note that the average ellipticity appears in this equation, and that the relation between the average ellipticity and the shear is generally non-linear.
We recall here that other ellipticity estimators can be defined in addition to the one presented  in Eq. ~(\ref{eq:ellipticity}). Another common estimator is
\begin{equation}
\epsilon=\frac{(Q_{11}-Q_{22})+2iQ_{12}}{Q_{11}+Q_{22}+2(Q_{11}Q_{22}-Q_{12}^2)^{1/2}}\;,
\end{equation}
which has a perfect response to shear, i.e. the shear responsivity is 1 \citep{Seitz97}.  However this estimator is considered more noisy and therefore not commonly used in weak-lensing measurements, and in particular it is not used by KSB. For this reason we will employ $\chi$ as ellipticity estimator rather than $\epsilon$ throughout this work.

\section{Shear measurements}

In practice, shear estimates are obtained from small and noisy background galaxies. The observed shape of any object is the result of a convolution of its intrinsic surface brightness $I^{0}(\vec{\theta})$ with the point spread function $P(\vec{\theta})$. The convolution tends to make the object more circular or to imprint a spurious ellipticity on it if the PSF is not isotropic. Moreover, any measurement of moments has to incorporate a weight function in order to suppress the pixel noise dominating at large spatial scales. Convolution and weighting change the surface brightness to
\begin{equation}
I^{obs}(\vec{\theta})=W(\vec{\theta})\int I^{0}(\vec{\theta}^{\prime})P(\vec{\theta}-\vec{\theta}^{\prime})d^2\theta^{\prime}.\label{eq:obsI}
\end{equation}
Since we are interested in the object's unconvolved and unweighted shape, we need to correct these two effects. In this Section, we first assume $P(\vec{\theta}-\vec{\theta}^{\prime}) \rightarrow \delta(\vec{\theta}-\vec{\theta}^{\prime})$, i.e.~we neglect the PSF convolution, and postpone the PSF correction to the following Section.

\subsection{Standard KSB}
We review in this section the standard KSB formalism, neglecting PSF convolution. In this situation the only complication is given by the presence of the weighting function for the computation of moments, which modifies the relation between shear and ellipticity given by Eq. ~(\ref{eq:chi_g}). 
Weighting changes Eq.~(\ref{eq:ellipticity}) to
\begin{equation}
\chi_{\alpha}=\frac{1}{Tr(Q)}\int d^2 \theta I^{obs}(\vec{\theta})\eta_{\alpha}W\left(\frac{|\vec{\theta}|^2}{\sigma^2}\right)\;,
\label{eq:chilens}
\end{equation} 
with 
\begin{equation}
\eta_{\alpha} = 
\left\{ 
\begin{array}{ccl} 
\theta_{1}^{2}-\theta_{2}^{2} & \mbox{if} & \alpha=1 \\ 
2\theta_1 \theta_2 & \mbox{if} & \alpha=2
\end{array} 
\right.\;.
\end{equation}
Note that also $TrQ$ in Eq. ~(\ref{eq:chilens}) is evaluated using weighted moments.
Using Eq.~(\ref{eq:coordtransf}) and the conservation of the surface brightness, $I^{obs}(\vec{\theta})=I^{s}(A\vec{\theta})$, we can infer the surface brightness in the source plane. From its second moments,
\begin{eqnarray}
Q^{s}_{ij}&=&\int d^2 \beta I^{s}(\vec{\beta})\beta_{i}\beta_{j}W\left(\frac{|\vec{\beta}|^2}{\hat{\sigma}^2}\right)\nonumber \\
&=&(\det{A})A_{ik}A_{il} \int d^2\theta I^{obs}(\vec{\theta})\theta_{k}\theta_{l}\nonumber \\
&\times& W\left(\frac{(|\vec{\theta}|^2-2\eta_{\alpha}g^{\alpha}+|\vec{\theta}|^2|g|^2)}{\sigma^2(1+|g|^2)}\right)\;,\label{eq:Qs}
\end{eqnarray}
we form the ellipticity
\begin{equation}
\chi^{s}_{\alpha}=C\int d^2 \theta I^{obs}(\vec{\theta})\xi_{\alpha}
W\left(\frac{|\vec{\theta}|^2-2\eta_{\beta}g^{\beta}+|\vec{\theta}|^2|g|^2}{\sigma^2(1+|g|^2)}\right)\;,
\label{eq:chisource}
\end{equation}
where 
\begin{equation}
\label{eq:CdetA}
 C = \frac{(\det A)(1-\kappa)^2}{Tr(Q^s)} \ \mathrm{and}
\end{equation}
\begin{equation}
\xi_{\alpha}=\eta_{\alpha}-2g_{\alpha}|\vec{\theta}|^2+(-1)^{\alpha}\eta_{\alpha}(g_1^2-g_2^2)+
2g_1g_2 \eta^{\dagger}_{\alpha}\;.
\end{equation}
The relation between the two filter scales in Eq. ~(\ref{eq:Qs}) is given by $\hat{\sigma}^2=(1-\kappa)^2(1+|g|^2)\sigma^2$ and the multiplicative term $(\det A)(1-\kappa)^2$ in Eq. ~(\ref{eq:CdetA}) will cancel out once $Tr(Q^{s})$ is written in terms of $Tr(Q)$.
Note that Einstein's sum convention is not implied in $(-1)^{\alpha}\eta_{\alpha}$, and that
\begin{equation}
\eta^{\dagger}_{\alpha} = 
\left\{ 
\begin{array}{ccl} 
\eta_2 & \mbox{if} & \alpha=1 \\ 
\eta_1 & \mbox{if} & \alpha=2
\end{array} 
\right.\;.
\end{equation}
We adopt this notation for a general tensor,
\begin{equation}
\Omega^{\dagger}_{\alpha \beta...\zeta} = 
\left\{ 
\begin{array}{ccl} 
\Omega_{2 \beta...\zeta} & \mbox{if} & \alpha=1 \\ 
\Omega_{1 \beta...\zeta} & \mbox{if} & \alpha=2
\end{array} 
\right.
\end{equation} 

Combining Eqs.~(\ref{eq:chilens}) and (\ref{eq:chisource}) gives a more complicated relation between ellipticities in the source and in the lens planes than Eq.~(\ref{eq:chi_g}) due to the presence of the weight function. Keeping only first-order terms in $g$, this relation is
\begin{equation}
\chi_{\alpha}-\chi_{\alpha}^{s}=g^{\beta}P^{sh}_{\alpha \beta},\label{Psh}
\end{equation}
\citep{KSB95,Hoekstra98} with
\begin{equation}
P^{sh}_{\alpha \beta}=-2\frac{\chi_{\alpha}L_{\beta}}{Tr(Q)}-2\chi_{\alpha}\chi_{\beta}+
2\frac{B_{\alpha \beta}}{Tr(Q)}+2\delta_{\alpha \beta}
\end{equation}
and
\begin{eqnarray}
L_{\beta}&=&\frac{1}{\sigma^2}\int d^2\theta I^{obs}(\vec{\theta})
W^{\prime}|\vec{\theta}|^2 \eta_{\beta}\;,\nonumber\\
B_{\alpha \beta}&=&\frac{1}{\sigma^2}\int d^2\theta I^{obs}(\vec{\theta})
W^{\prime}\eta_{\alpha} \eta_{\beta}\;.
\label{eq:defBL}
\end{eqnarray}
The notation we use here follows \cite{Bartelmann01}. 

\subsection{Shear estimates}

Equation~(\ref{Psh}) directly relates the measured weighted ellipticity $\chi$ to the shear $g$ if the intrinsic ellipticity of the source $\chi^s$ is known. Since $\chi$ and $\chi^s$ cannot be disentangled for individual galaxies, averages over ensembles of images are necessary to estimate $g$,
\begin{equation}
\langle \chi_{\alpha} \rangle -\langle \chi^{s}_{\alpha} \rangle = \langle g^{\beta}P^{sh}_{\alpha \beta} \rangle \rightarrow \langle g_{\alpha} \rangle = g_{\alpha} = \langle P^{sh} \rangle ^{-1}_{\alpha \beta}\langle \chi^{\beta} \rangle\;.
\end{equation}
However, the quantity which is commonly computed is
\begin{equation}
\langle \tilde{g}_{\alpha} \rangle =\langle (P^{sh} )^{-1}_{\alpha \beta}\chi^{\beta} \rangle\;,
\end{equation}
\cite{Erben01}, assuming that $\langle (P^{sh})^{-1}\chi^{s} \rangle =0$. This condition is not guaranteed since $P^{sh}$ itself depends on $\chi$. The symbol $\tilde{g}_{\alpha}$ denotes the shear estimate obtained by solving Eq.~(\ref{Psh}) with $\chi^{s}=0$. We introduce it since $\tilde{g}$ is not the true shear (which is inaccesible for a single galaxy), but the shear one would measure if the source was circular. The true shear $g$ is then sought by averaging $\tilde{g}$. Equation~(\ref{eq:chi_g}) shows that for $\chi^{s}=0$ and $W(x)=1$, $\langle \tilde{g} \rangle$ is related to $\chi$ by
\begin{equation}
\langle \tilde{g} \rangle = \left \langle \frac{1-\sqrt{1-\chi^2}}{\chi}\right \rangle \simeq \left\langle  \frac{\chi}{2} + \frac{ \chi^3}{8}+ \frac{\chi^5}{16} + ... \right \rangle\;.
\label{eq:gOK}
\end{equation}
In general, $\langle \tilde{g} \rangle$ differs from the true shear $g$ computed in Eq.~(\ref{eq:gave}). 
Assuming that $g \ll 1$, meaning $\langle \chi \rangle \ll 1$, and the distribution of the intrinsic ellipticities to be Gaussian with standard deviation $\sigma_{\chi}$, the difference can be written as: 
\begin{equation}
g-\langle \tilde{g} \rangle \simeq \frac{\langle \chi \rangle}{2}\left(\frac{\sigma^2_{\chi}}{1-\sigma^2_{\chi}}\right)-\frac{3\sigma^2_{\chi}\langle \chi \rangle}{8}
\end{equation}
from which
\begin{equation}
\langle \tilde{g} \rangle \simeq g\left(1-\frac{1}{4}\sigma^2_{\chi}\right)\;.
\end{equation}
For a realistic $\sigma_{\chi} \simeq 0.3$, the bias introduced by averaging shear estimates instead of ellipticities is $\approx2 \%$.

Moreover, averaging shear estimates does not allow one to assume that $\tilde{g}$ is small, as done in the original derivation of $P^{sh}$, since it is always of the same magnitude as $\chi$. In coordinates rotated such that $\tilde{g}$ has only one non-vanishing component, and in absence of a weight function, the relation between $\chi$ and $\tilde{g}$ provided by KSB in Eq.~(\ref{Psh}) is
\begin{equation}
\tilde{g}^{KSB} \simeq \frac{\chi_1}{2}+\frac{\chi_1^3}{2}+\frac{\chi_1^5}{2}+...\;.
\label{eq:gKSB}
\end{equation} 
Obviously, this is correct only to lowest order. Comparing Eqs. ~(\ref{eq:gOK}) and ~(\ref{eq:gKSB}), the error made by KSB in the shear estimation is a function of the measured ellipticity and scales as $(3\chi^3/8+7\chi^5/16)$. Typically, $|\chi| \in[0.5...0.8]$, implying that the bias KSB introduces in the shear estimate (without weight function) is in the range $[6...33] \%$. The reason for this bias comes from the fact that second- or higher-order terms in $g$ have been neglected in the derivation of Eq.~(\ref{Psh}), while terms like $\chi^2 g$ have been kept.  Once $g$ is identified with $\tilde{g}$, these mixed terms are effectively of the same order as the $g^3$ terms. In a consistent first-order relation between $\chi$ and $\tilde{g}$, only the first-order term in $\chi g$ can be considered. Then, $P^{sh}_{\alpha \beta}$ looks like
\begin{equation}
P^{sh,(0)}_{\alpha \beta}=\frac{2B_{\alpha \beta}}{TrQ}+2\delta_{\alpha \beta}\;.\label{eq:Psh0}
\end{equation}
We shall refer to this approximation as KSB1. In this case, the solution for $\tilde{g}$ is
\begin{equation}
\tilde{g}^{KSB1}= \frac{\chi}{2}\;.\label{eq:approxKSB1}
\end{equation}
The error on the shear estimate made by KSB1 scales like $(-\chi^3/8-\chi^5/16)$, leading to an underestimate which is considerably smaller than the overestimate given by KSB. However, as discussed before, $\chi$ is practically never small, meaning that first-order approximations may be poor.

In a frequently used variant of KSB, $P^{sh}_{\alpha \beta}$ is approximated by half its trace (KSBtr hereafter),
\begin{equation}
P^{sh}_{\alpha \beta} \simeq \frac{1}{2}Tr(P^{sh}_{\alpha \beta})\delta_{\alpha \beta}\;.
\end{equation}
This is usually justified saying that the trace is less noisy than the inverse of the full tensor, as we shall show in Sect.~3.3. This statement is certainly correct for large ellipticities. However, it turns out to work much better than the full tensor even in the absence of noise, PSF and weighting. The reason is that it leads to the relation
\begin{equation}
\tilde{g}^{KSBtr} \simeq \frac{\chi}{2}+\frac{\chi^3}{4}+\frac{\chi^5}{8}+...\label{eq:approxKSBtr}
\end{equation}
between $\tilde{g}$ and $\chi$, which biases the shear estimate by $\chi^3/8+\chi^5/16$.
 
We can summarise the preceding discussion as follows:
\begin{enumerate}
\item KSB incorrectly approximates Eq.~(\ref{eq:chi_g});
\item KSB1 is mathematically consistent;
\item KSBtr approximates Eq.~(\ref{eq:chi_g}) better even though it lacks mathematical justification;
\item No KSB variant discussed so far is correct to third order in $\chi$.
\end{enumerate}

\subsection{Third-order relation between $g$ and $\chi$}

We now derive a consistent third-order relation between $\chi$ and $\tilde{g}$, including the effects of the weight function. We follow closely the approach in Sect.~4.6.2 of \cite{Bartelmann01}, and use Einstein's sum convention. We start from Eq.~(\ref{eq:chisource}) and Taylor-expand the weight function around $g=0$ to third order in $g$,
\begin{equation}
\begin{split}
&W\left(\frac{(|\vec{\theta}|^2-2\eta_{\alpha}g^{\alpha}+|\vec{\theta}|^2|g|^2)}{\sigma^2(1+|g|^2)}\right) \simeq  \\
&W\left(\frac{|\vec{\theta}|^2}{\sigma^2}\right)-2W^{\prime}\left(\frac{|\vec{\theta}|^2}{\sigma^2}\right)\frac{\eta_{\beta}g^{\beta}(1-|g|^2)}{\sigma^2}+  \\ 
&2W^{\prime \prime}\left(\frac{|\vec{\theta}|^2}{\sigma^2}\right)\frac{(\eta_{\beta}g^{\beta})^2}{\sigma^4}-
 \frac{4}{3}W^{\prime \prime \prime}\left(\frac{|\vec{\theta}|^2}{\sigma^2}\right) \frac{(\eta_{\beta}g^{\beta})^3}{\sigma^6}+ \mathcal{O}(g^{4}),
\end{split}
\end{equation}
where
\begin{equation}
\frac{\eta_{\beta} g^{\beta}}{(1+g^2)}\simeq \eta_{\beta}g^{\beta}(1-g^2) + \mathcal{O}(g^{4})
\end{equation}
was used. Note that the derivatives of the weight function are taken with respect to $\theta^2$. Truncating the series at a given order implies that the final result will depend on the shape of the weight function.

We proceed with the calculation of $\chi^{s}_{\alpha}Tr(Q^{s})$ to third order in $\chi g$,
\begin{equation}
\begin{split}
&\frac{\chi^{s}_{\alpha}Tr(Q^{s})}{(\det A)(1-\kappa)^2}=\int d^2\theta \xi_{\alpha}I(\vec{\theta})W\left(\frac{|\vec{\theta}|^2-2\eta_{\beta}g^{\beta} + |\vec{\theta}|^2g^2}{\sigma^2}\right)= 
\\ 
& \chi_{\alpha}Tr(Q)-2g^{\beta}B_{\alpha \beta}+2g^{\beta}g^{\gamma}D_{\alpha \beta \gamma}-2g_{\alpha}Tr(Q)+4g_{\alpha}g^{\beta}L_{\beta}+
\\
&-4K_{\beta \gamma}g_{\alpha}g^{\beta}g^{\gamma}+(-1)^{\alpha}(g_2^2-g_1^2)\chi_{\alpha}Tr(Q)
\\
&-2(-1)^{\alpha}B_{\alpha \beta}(g_2^2-g_1^2)g^{\beta}+2g_1g_2\chi_{\alpha}^{\dagger} Tr(Q)+
\\
& -4B^{\dagger}_{\alpha \beta}g^{\beta}g_{1}g_{2}-\frac{4}{3}U_{\alpha \beta \gamma \delta}g^{\beta}g^{\gamma}g^{\delta}+\mathcal{O}(g^{4}),
\end{split}
\end{equation}
where the definitions
\begin{eqnarray}
D_{\alpha \beta \gamma}&=&\frac{1}{\sigma^4}\int d^2\theta I^{obs}(\vec{\theta})
W^{\prime \prime}\eta_{\alpha}\eta_{\beta}\eta_{\gamma}\;,\nonumber\\
U_{\alpha \beta \gamma \delta}&=&\frac{1}{\sigma^6}\int d^2\theta I^{obs}(\vec{\theta})
W^{\prime \prime \prime}\eta_{\alpha}\eta_{\beta}\eta_{\gamma}\eta_{\delta}
\end{eqnarray}
appear. $L_{\alpha}$ and $B_{\alpha \beta}$ are given in Eq. \ref{eq:defBL}. In the same way, we evaluate
\begin{equation}
\begin{split}
&\frac{Tr(Q^{s})}{(\det A)(1-\kappa)^2} \simeq Tr(Q)(1+|g|^2) -2g^{\alpha}L_{\alpha}+2g^{\alpha}g^{\beta}K_{\alpha \beta}-  \\
&2g^{\alpha}\chi_{\alpha}Tr(Q)+4g^{\alpha}g^{\beta}B_{\alpha \beta}-4D_{\alpha \beta  \gamma}g^{\alpha}g^{\beta}g^{\gamma}-\frac{4}{3}J_{\alpha \beta \gamma}g^{\alpha}g^{\beta}g^{\gamma} \\ 
&=Tr(Q)(1+f(g)),
\end{split}
\end{equation}
where we implicity defined $f(g)$ and
\begin{eqnarray}
K_{\alpha \beta}&=&\frac{1}{\sigma^4}\int d^2\theta I^{obs}(\vec{\theta})
W^{\prime \prime}|\vec{\theta}|^2\eta_{\alpha} \eta_{\beta}\;,\nonumber\\
J_{\alpha \beta \gamma}&=&\frac{1}{\sigma^6}\int d^2\theta I^{obs}(\vec{\theta})
W^{\prime \prime \prime}|\theta|^2 \eta_{\alpha}\eta_{\beta}\eta_{\gamma}\;.
\end{eqnarray}
From these quantities, we compute
\begin{equation}
\chi_{\alpha}-\chi^{s}_{\alpha}=\frac{\chi_{\alpha}Tr(Q)(1+f(g))-\chi^{s}_{\alpha}Tr(Q^{s})}{Tr(Q)(1+f(g))}.\label{eq:approx2}
\end{equation}
This equation holds exactly in absence of a weight function. If a weight function is included, $f(g)$ is at most of order $0.02$, and we shall consider Eq.~(\ref{eq:approx2}) exact to third order. After some algebra we find
\begin{equation}
\chi_{\alpha}-\chi_{\alpha}^{s}=\frac{g^{\beta}[P_{\alpha \beta}+g^{\gamma}(R_{\alpha \beta \gamma}+g^{\delta}S_{\alpha \beta \gamma \delta})] + \Xi_{\alpha}+\mathcal{O}(g^{4})}{1+f(g)}\label{eq:chi_sh}
\end{equation}
where
\begin{eqnarray}
R_{\alpha \beta \gamma}&=&2\frac{\chi_{\alpha}K_{\gamma \beta}}{Tr(Q)}+
4\frac{\chi_{\alpha}B_{\gamma \beta}}{Tr(Q)}-2\frac{D_{\alpha \beta \gamma}}{Tr(Q)}-
4\frac{\delta_{\alpha \gamma}L_{\beta}}{Tr(Q)}\;,\nonumber\\
S_{\alpha \beta \gamma \lambda}&=&\frac{2K_{\beta \gamma}\delta_{\alpha \lambda}}{TrQ}+
\frac{4}{3}\frac{U_{\alpha \beta \gamma \lambda}}{TrQ}\;,
\end{eqnarray}
and
\begin{equation}
\begin{split}
&\Xi_{\alpha}=\left(\chi_{\alpha}-\frac{2B_{\alpha \beta}g^{\beta}}{TrQ}\right)|g|^2 -(-1)^{\alpha}(g_2^2-g^2_1)\left(\chi_{\alpha}-\frac{2B_{\alpha \beta}g^{\beta}}{TrQ}\right)- \\
&2g_1g_2\left(\chi_{\alpha}^{\dagger}-\frac{2B^{\dagger}_{\alpha \beta}g^{\beta}}{TrQ}\right)
\end{split}
\end{equation}
Introducing second and third-order terms leads to a non-linear relation between $\chi$ and $g$ which needs to be solved numerically. Moreover, sixth-order moments of the observed surface-brightness distribution appear in $R_{\alpha \beta \gamma}$ and eigth-order moments in $U_{\alpha \beta \gamma \delta}$ because of the Taylor expansion of the weight function to third order. We discuss in the following Section how to deal with the non-linear relation between shear and ellipticity and possible noise issues due to the appearance of higher moments.

\begin{figure*}
 \begin{minipage}{170mm}
 	\centerline{
 	\psfig{figure=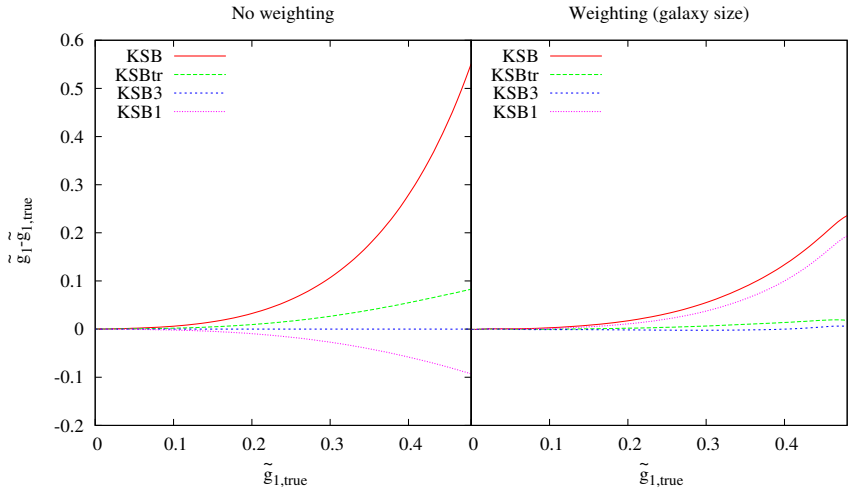,width=15cm,angle=270}}
\caption{Shear estimate $\tilde{g}_1$ as a function of the applied shear for noise-free and unconvolved S\'ersic-type galaxy images as provided by KSB (red line), KSBtr (green line), KSB3 (blue line), and KSB1 (magenta line). In the \textit{left panel} no weighting function has been used to measure moments of the light distribution, while in the \textit{right panel} a Gaussian weighting function has been employed with a width equal to the size of the object. The effective galaxy radius was $R_e=2$ pixel, the S\'ersic model was tenfold oversampled, and the image sidelength was $40$ pixels.}\label{fig:noLin}
\end{minipage}
\end{figure*}

\subsection{Tests}

We now show the results of simple tests carried out to check how well the four variants of KSB estimate the shear. We consider a circular source ($\chi^{s}=0$) with a S\'ersic brightness profile,
\begin{equation}
I(r)=I_{0}\exp \left[-b_{n_s}\left(\left(\frac{r}{R_e}\right)^{1/n_s}-1\right)\right]\label{eq:Sersic}
\end{equation}
where $R_e$ is the radius containing half of the flux and $n_s$ the S\'ersic index and $b_{n_s}$ is a constant which depends on $n_s$. This type of profile is identical to a Gaussian for $n_s = 0.5$ and is steeper in the centre for $n_s > 0.5$. In the following test, we assume $n_s=1.5$, which represents the average value for rather bright galaxies in the COSMOS field \citep{Sargent07}. We shear this profile by a variable amount $\tilde{g}_1$, keeping  $\tilde{g}_2=0.1$ fixed, using Eq.~(\ref{eq:coordtransf}). For all following tests, the effective galaxy radius was $R_e = 2$ pixels, the S\'ersic model was tenfold oversampled, and the image sidelength was set such as to not truncate the galaxy at
the image boundary. Then, we measure the ellipticity as defined in Eq.~(\ref{eq:ellipticity}). Since the model galaxy is intrinsically circular, the source ellipticity is entirely generated by the applied shear which is varied in a wide range such as to mimic the intrinsic ellipticity dispersion. The weight function has been chosen as Gaussian with $\sigma = 2R_e$. We repeated this test assuming a flat weight function ($W(x)=1$) in order to estimate how much the different approximations in deriving $P^{sh}$ affect the measurement. The results are shown in Fig.~\ref{fig:noLin}.

In absence of a weight function (left panel of Fig.~\ref{fig:noLin}), the performance of the four variants closely follows the analytic behaviour worked out in Sects.~3.1 and 3.2: KSB severely overestimates the shear for large $\tilde{g}_1$, while KSBtr and KSB1 better approximate the shear. KSB3 returns the correct shear under this condition.  

The weight function renders the image more circular and thus reduces the measured $\chi$. This means that the high-order terms in $\chi$ contribute less to the shear estimate. Therefore, the deviation from the correct result is significantly lower for all the methods (right panel of Fig.~\ref{fig:noLin}). This is not true for KSB1, which allows only a first-order correction for the weight function.

\begin{figure}
	\centerline{
       \psfig{figure=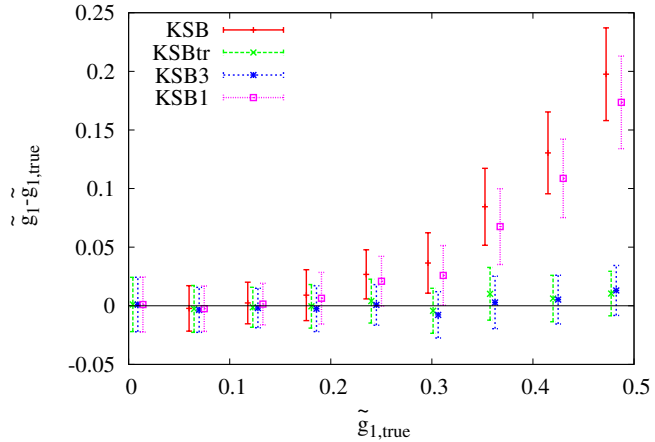,width=9cm,angle=270}}
\caption{Shear estimate $\tilde{g}_1$ as a function of the applied shear for noisy but unconvolved S\'ersic-type galaxy images as provided by KSB (red line), KSBtr (green line), and KSB3 (blue line). The total flux of the source was fixed to unity, the noise rms to $10^{-3}$. The average is taken over 100 objects. Errorbars denote standard deviation of the mean.} \label{fig:Noise}
\end{figure}

We also investigate the behaviour of the four KSB variants for realistic pixel noise. The average result for 100 galaxies is shown in Fig.~\ref{fig:Noise}.

KSBtr is the only method for which no matrix inversion is required. It is thus not surprising that it exhibits the lowest standard deviation for all values of $\tilde{g}$. KSB and KSB3 have a comparable amount of noise even though KSB3 involves the computation of 6th and 8th moments of the light distribution. The reason is that these higher-order moments  are computed using the second and third derivatives of the weight function. There is no price to be payed (in terms of measurement noise) in using KSB3 instead of the simple KSB description.
\begin{figure*}
 \begin{minipage}{170mm}
 	\centerline{
 	\psfig{figure=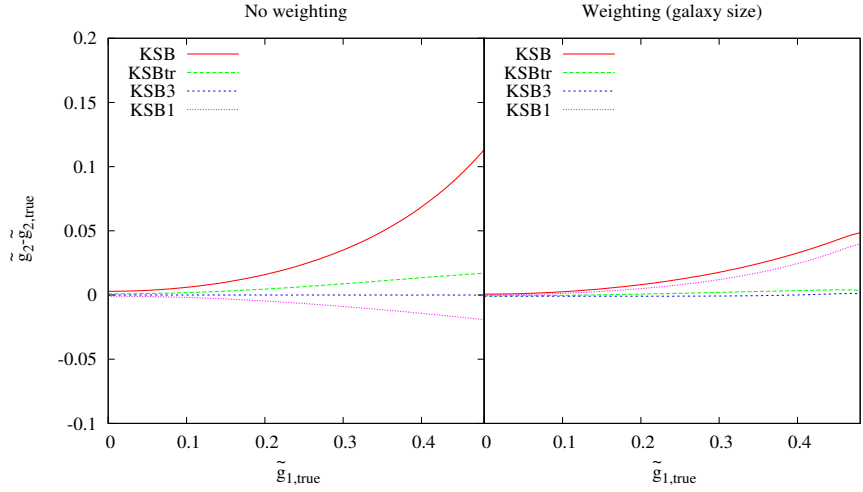,width=15cm,angle=270}}
\caption{Shear estimation cross-talk for $\tilde{g}_2$ as a function of the applied shear $\tilde{g}_1$ for noise-free and unconvolved S\'ersic-type galaxy images as provided by KSB (red line), KSBtr (green line), KSB3 (blue line), and KSB1 (magenta line). In the \textit{left panel} no weighting function has been used to measure moments of the light distribution, while in the \textit{right panel} a Gaussian weighting function has been employed with a width equal to the size of the object.}\label{fig:CrossTalk}
\end{minipage}
\end{figure*}
We also investigate how much the measurement of one component of the shear is affected by the value of the other component. For this case, we also studied the case of unweighted and weighted moment measurements. The result is shown in Fig.~\ref{fig:CrossTalk}. The obvious cross-talk between the two components is not surprising for KSB, KSBtr or KSB1 since all terms which mix $\tilde{g}_1$ and $\tilde{g}_2$ were neglected in the calculation. Introducing third-order corrections, the estimate of one shear component becomes almost independent of the other component.

\begin{figure}
	\centerline{
      \psfig{figure=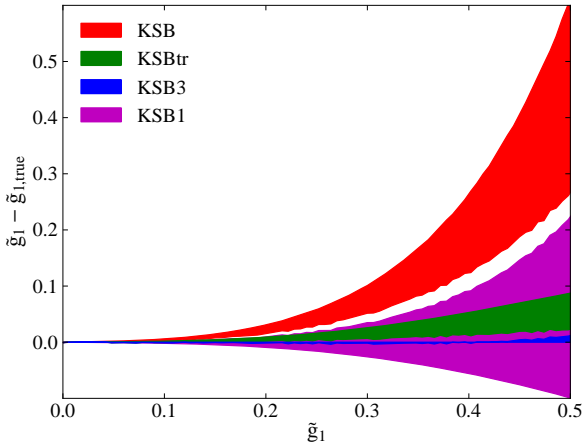,width=8cm,angle=0}}
\caption{Dependence of the shear estimate $\tilde{g}_1$ on the size of the weighting function width as a function of the applied shear for noise-free and unconvolved S\'ersic-type galaxy images as provided by KSB (red), KSBtr (green), KSB3 (blue), and KSB1 (magenta). We consider a Gaussian weighting function with width $\sigma = [2R_e,...,\infty]$. The lower limits correspond to $\sigma=2R_e$, and the upper limits are identical to the unweighted case shown in Fig. \ref{fig:noLin}.}
\label{fig:weightSize}
\end{figure}

Finally, we study how much the bias in the shear measurement depends on the width $\sigma$ of the weight function $W$. We vary the width within $[2R_e, \infty)$. The result is shown in Fig.~\ref{fig:weightSize}. KSB and KSB1 exhibit a strong dependence on $\sigma$, while KSBtr is more robust, and KSB3 is almost independent of $\sigma$. Due to the poor correction of the weight-induced change of $\chi$, KSB1 performs poorest in this test. For KSB, the reduction of $\chi$ due to the weighting limits its strong non-linear response such that the bias decreases for narrow weight functions. As KSB3 employs the best description of the weighting, it performs excellently in this test.

\begin{figure}
	\centerline{
       \psfig{figure=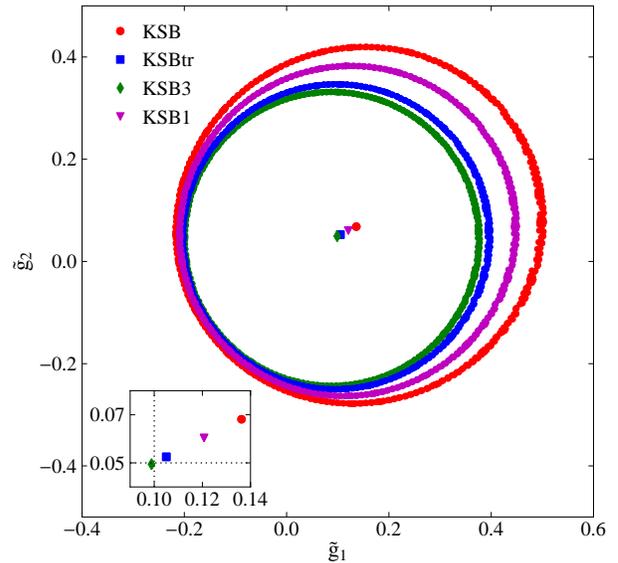,width=8cm,angle=0}}
\caption{Shear estimates of a sample of S\'ersic-type galaxies with $\chi^s = 0.3$ after shear $g=(0.1,0.05)$ is applied. Red dots are the results from KSB, magenta dots from KSB1, blue dots from KSBtr, and green dots from KSB3. The dots in the center show the position of the ensemble averages of the estimates. A zoom of the central region is shown in the small panel, where the intersection of the dotted lines indicates the outcome of a perfect measurement.}\label{fig:ringTest}
\end{figure}

In all tests carried out so far, we have assumed that the intrinsic ellipticity of the object vanishes, $\chi^s=0$. This is of course idealised since galaxies have an intrinsic ellipticity dispersion. In order to test the performance of the four methods for an isotropic source-ellipticity distribution, we apply the so-called \textit{ring test} \citep{Nakajima07}. We construct an ensemble of test galaxies falling on a circle in the ellipticity plane, shear them, measure their shapes, and take the mean. We choose an intrinsic ellipticity $|\chi^{s}|=0.3$ and apply the shear $g=(0.1,0.05)$. The result is shown in Fig.~\ref{fig:ringTest}. A perfect method would recover the correct shear after averaging over all test galaxies. Not surprisingly, we find that KSB is unable to recover the correct shear from the averaged individual shear estimates since they depend non-linearly on $\chi$. This leads to an average overestimate of $\approx35\,\%$ if the shear is aligned with the intrinsic ellipticity. As the other variants have lower non-linear error in the $\chi$-$g$ relation, the mean values are biased by $\approx20\,\%$ (KSB1), $\approx5\,\%$ (KSBtr) and $\approx1\,\%$ (KSB3).

\section{PSF convolution}

Any measured galaxy's ellipticity is the result of three distinct physical processes: intrinsic ellipticity, lensing, and PSF convolution. As discussed above, lensing maps the galaxy's light distribution from the source to the lens plane, distorting its shape. The relation between galaxy ellipticity and shear can be found solving Eq.~(\ref{eq:chi_g}) if there is no weight function, or Eq.~(\ref{eq:chi_sh}) if weighted moments are used to define the ellipticity. In general, the equation one needs to solve to relate ellipticity to a shear estimator $\tilde{g}$ has the implicit form
\begin{equation}
\chi=f(\tilde{g},\chi)\;.
\end{equation}
On the other hand, PSF deconvolution maps the observed ellipticity from the image plane (on which the object is lensed and convolved with the PSF) to the lens plane (on which the object is lensed only)
\begin{equation}
\chi = h(\chi^{obs}) \;.
\end{equation} 
Thus, the relation between observed ellipticity and the shear estimator in presence of PSF convolution is the solution of
\begin{equation}
\chi^{obs}=h^{-1}[f(\tilde{g},h(\chi^{obs}))]\;.\label{eq:okmapping}
\end{equation}

If the PSF is perfectly circular, the only effect of $h$ is a circularisation of the object, otherwise the PSF induces additional anisotropic distortions. Therefore, it is crucial to properly correct these two effects in order to realiably estimate the shear. We consider in the following the case of a spherical PSF and briefly discuss the case of an anisotropic PSF in Sect. \ref{sec:psf_anisotropy}.

Instead of carrying out a proper PSF deconvolution first and then estimating the shear using the unconvolved ellipticity, as summarised by Eq.~(\ref{eq:okmapping}), KSB links the observed ellipticity to the shear by the following approach:
\begin{equation}
\chi^{obs}_{\alpha}=\chi^{sh}_{\alpha}-\chi^{g}_{\alpha},\label{eq:full}
\end{equation}
where $\chi^{sh}_{\alpha}$ is given by Eq.~(\ref{eq:chi_sh}), and $\chi^{g}$ is
\begin{equation}
\chi^{g}_{\alpha} = P^{sm}_{\alpha \gamma}(P^{sm,*})^{-1}_{\gamma \beta}\chi^{sh,*}_{\beta}.
\label{eq:chi_g_f}
\end{equation}
$P^{sm}$ is the so-called "smear polarisability tensor" and has the form
\begin{equation}
P^{sm}_{\alpha \beta}=\frac{1}{Tr(Q)}\left[\left(M+\frac{2Tr(Q^{\prime})}{\sigma^2}\right)\delta_{\alpha \beta}+G_{\alpha \beta}-\chi_{\alpha}(2F_{\beta}+L^{\prime}_{\beta})\right],
\end{equation}
where 
\begin{eqnarray}
M&=&\int d^2\theta I(\vec{\theta})W\left(\frac{|\vec{\theta}|^2}{\sigma^2}\right),\nonumber\\
F_{\alpha}&=&\frac{1}{\sigma^2}\int d^2\theta I(\vec{\theta})W^{\prime}\left(\frac{|\vec{\theta}|^2}{\sigma^2}\right)\eta_{\alpha},\quad\mbox{and}\nonumber\\
G_{\alpha \beta}&=&\frac{1}{\sigma^4}\int d^2\theta I(\vec{\theta})W^{\prime \prime}\left(\frac{|\vec{\theta}|^2}{\sigma^2}\right)\eta_{\alpha}\eta_{\beta}.
\end{eqnarray}
$L^{\prime}_{\alpha}$ has to be interpreted as $L_{\alpha}$ calculated with the second derivative of the weight function, while $Tr(Q^{\prime})$ and $Tr(Q)$ are calculated with the first derivative of the weight function. We refer to Sect.~4.6.2 of \cite{Bartelmann01} for a complete derivation of Eq.~\ref{eq:full}.

Since $\chi^{sm}$ encodes the action of lensing (cf. last Section on the appropriate forms of this mapping), we can rewrite Eq.~\ref{eq:full} as
\begin{equation}
\chi^{obs}=f(\tilde{g},\chi^{obs})-\chi^{g}(\tilde{g},\chi^{obs}).\label{eq:ksbmapping}
\end{equation}
It is important to note that the lensing-induced mapping is now evaluated with the observed, i.e. convolved, ellipticity instead of the unconvolved ellipticity. This approach therefore requires the correction term $\chi^g$, which corresponds to a correct treatment of the PSF convolution (Eq.~(\ref{eq:okmapping})) if and only if
\begin{equation}
\chi^{g}(\tilde{g},\chi^{obs})=f(\tilde{g},\chi^{obs})-h^{-1}[f(\tilde{g},h(\chi^{obs})].\label{eq:mapping}
\end{equation}

\begin{figure}
 	\centerline{
       \psfig{figure=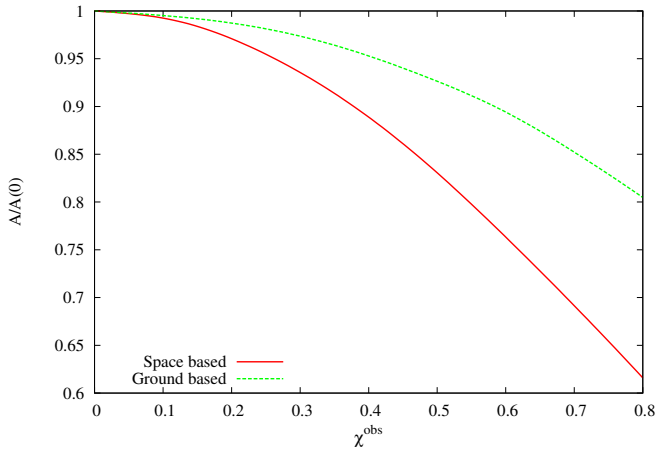,width=9cm,angle=270}}
\caption{Dependence of $A$ as defined in Eq.~(\ref{eq:A}) on the observed ellipticity. Red line represents the case of PSF with $\mathrm{FWHM}=0.5R_e$ and $\beta=2$ to mimic a space-based observation and the green line the case of a PSF with $\mathrm{FWHM}=5R_e$ and $\beta=5$ to mimic a ground-based observation. In the case of a infintely wide PSF we would have $A/A(0)=1$ independently on the ellipticity.}\label{fig:Achi}
\end{figure}

We study now a very simple but instructive case. We assume a perfectly circular source, no weight function, an isotropic PSF, and shear oriented in a single direction. Then, $P^{sm}_{\alpha \beta}$ becomes diagonal,
\begin{equation}
P^{sm}_{\alpha \beta}=\frac{M}{TrQ}\delta_{\alpha \beta}\;.
\end{equation}
In a forthcoming paper (Melchior et al., in prep.) we shall demonstrate how to do a proper PSF deconvolution, using the moments of the PSF and the convolved object, and show that the mapping $h$ between the convolved ellipticity $\chi^{obs}$ and the unconvolved ellipticity $\tilde{\chi}$ in the lens plane is given by
\begin{equation}
h(\chi^{obs})= \tilde{\chi}= \frac{\chi^{obs}}{1-A(\chi^{obs})},
\end{equation}
where
\begin{equation}
A=\frac{M}{TrQ}\frac{TrQ^{*}}{M^{*}}\label{eq:A}
\end{equation}
is a functionof the observed ellipticity (as shown in Fig.~\ref{fig:Achi}) and of the size of the PSF (as shown in Fig.~\ref{fig:APSF}), and is bound to $[0,1]$ .
If the shear has a single component and there is no weight function involved in the measurement, $f(\tilde{g},\chi^{obs})$ is
\begin{equation}
f(\tilde{g},\chi^{obs})=\frac{2\tilde{g}-2(\chi^{obs})^{2}\tilde{g}}{1+\tilde{g}^2-2\tilde{g}\chi^{obs}}\label{eq:fgchi}
\end{equation}
According to Eq.~(\ref{eq:chi_g_f}), in the KSB formalism $\chi^{g}$ has the form:
\begin{equation}
\chi^{g}(\tilde{g},\chi^{obs})=A(\chi^{obs})f(\tilde{g},0).\label{eq:kKsb}
\end{equation}
In particular, in standard KSB, $\chi^{g}(\tilde{g},\chi^{obs})=2\tilde{g}A(\chi^{obs})$.

Substituting this expression for $\chi^g$ in the lhs of  Eqs.~(\ref{eq:mapping}), we can conclude that KSB gives a proper description of PSF deconvolution only if the function $f(\tilde{g},\chi^{obs})$ can be decomposed into a product of two functions, one depending on $\tilde{g}$ only and one on $\chi^{obs}$ only. This is by no means guaranteed. A detailed analysis reveals that there are two limiting cases in which Eqs.~(\ref{eq:mapping}) holds: 
\begin{itemize}
\item The PSF width vanishes: 

$A(\chi^{obs})=\chi^g(\chi^{obs})=0 \Rightarrow \chi^{obs} = \chi$. 

\item The observed ellipticity vanishes.
\end{itemize}

While the first case is trivial (but irrelevant), the second case can only be realized -- for any finite PSF width -- by a conspiracy of intrinsic and lensing-induced ellipticity.

To study in detail the error commited by KSB in the attempts to correct for the PSF convolution, we solve Eq.~(\ref{eq:full}) explicitly, employing the four variants $\chi^{sh}$ of mapping $\chi$ onto $\tilde{g}$ presented in the previous Section,
\begin{eqnarray}
\tilde{g}^{KSB} &\simeq & \frac{\tilde{\chi}_{0}}{2}+\frac{\tilde{\chi}^{2}_{0}}{2}A^{\prime}(0)+ \label{eq:gchiA}\\
&+& \frac{\tilde{\chi}_{0}^3}{2}[(1-A(0))(1+A^{\prime \prime}(0)/2)+A^{\prime}(0)^2]\nonumber \\
&+& \mathcal{O}(\tilde{\chi}_{0}^{4}) \nonumber \\
\tilde{g}^{KSB1} &=& \frac{\tilde{\chi}}{2} \nonumber \\
\tilde{g}^{KSBtr} &\simeq & \frac{\tilde{\chi}_{0}}{2}+\frac{\tilde{\chi}^{2}_{0}}{2}A^{\prime}(0)+ \nonumber \\
&+& \frac{\tilde{\chi}_{0}^3}{4}[(1-A(0))(1+A^{\prime \prime}(0))+2A^{\prime}(0)^2]\nonumber \\
&+& \mathcal{O}(\tilde{\chi}_{0}^{4}) \nonumber \\
\tilde{g}^{KSB3} &\simeq & \frac{\tilde{\chi}_{0}}{2}+\frac{\tilde{\chi}^{2}_{0}}{2}A^{\prime}(0)+ \nonumber \\
&+& \frac{\tilde{\chi}_{0}^3}{8}[1+4A^{\prime}(0)^{2}+2A^{\prime \prime}(0)-2A(0)(2+A^{\prime \prime}(0))]\nonumber \\
&+& \mathcal{O}(\tilde{\chi}_{0}^{4})  \nonumber 
\end{eqnarray}
where
\begin{equation}
\tilde{\chi}_{0} \equiv \frac{\chi^{obs}}{1-A(0)}\label{eq:gchitilde}
\end{equation}
and $A^{\prime}(0)$ and $A^{\prime \prime}(0)$ are the first and the second derivatives of $A(\chi^{obs})$ computed for $\chi^{obs}=0$.
If the PSF correction works perfectly, the relation between $\tilde{\chi}$ and $\tilde{g}$ has the same form as the exact unconvolved solution of Eq.~(\ref{eq:gave}),
\begin{equation}
\tilde{g} \simeq \frac{\tilde{\chi}}{2}+\frac{\tilde{\chi}^3}{8} + \mathcal{O}(\tilde{\chi}^{5}). \label{eq:tildeg}
\end{equation}
We note first of all that Eq.~(\ref{eq:gchiA}) is written in terms of $\tilde{\chi}_{0}$, while Eq.~(\ref{eq:tildeg}) is written in terms of $\tilde{\chi}$, meaning that in general the solutions are different already at first order. However the error at first order ($\chi^{obs} \ll 1$) is mostly of order $~10^{-4}$ and therefore negligible. In the limit of a very wide PSF $A(\chi) \simeq A(0)$ we find the deviations from the exact solution $b=\tilde{g}-\tilde{g}^{KSB...}$,
\begin{equation}
  \label{eq:bias}
\begin{split}
b^{KSB} &= \frac{3-4A}{8}\tilde{\chi}^3 +\mathcal{O}(\tilde{\chi}^{5}) \\ 
b^{KSB1} &= -\frac{\tilde{\chi}^3}{8} + \mathcal{O}(\tilde{\chi}^{5})\\
b^{KSBtr} &= \frac{1-2A}{8}\tilde{\chi}^3 + \mathcal{O}(\tilde{\chi}^{5})\\
b^{KSB3} &= -\frac{A}{2}\tilde{\chi}^3 + \mathcal{O}(\tilde{\chi}^{5})
\end{split}
\end{equation}
It is worth noting that the PSF correction introduces a bias with preferred direction: Shear estimates decrease as the PSF width increases.

\subsection{PSF anisotropy}
\label{sec:psf_anisotropy}
An anisotropic PSF introduces spurious ellipticity in the image plane which must be corrected. The appropriate correction in KSB relies on the hypothesis that the PSF can be considered almost isotropic. This enables its decomposition into an isotropic part $P^{iso}$ and an anisotropic part $q$,
\begin{equation}
P(\vec{\theta})=\int d^2\phi q(\vec{\phi})P^{iso}(\vec{\theta}-\vec{\phi})\label{eq:PSFsplit}
\end{equation} 
Even this decomposition can be problematic for certain PSFs \citep{KK99}. For example, a PSF given by the sum of two Gaussians with constant ellipticity does not fulfill the equation above. Assuming that Eq.~(\ref{eq:PSFsplit}) is valid, one can find a relation, valid to first order in $q$, between the observed and the isotropic ellipticity,
\begin{equation}
\chi_{\alpha}^{iso}=\chi_{\alpha}^{obs}-(P^{sm}_{\alpha \beta})q^{\beta}\label{eq:chiisoobs}
\end{equation}
The term $q_{\alpha}$, carrying information on anisotropies in the PSF, can be determined from the shape of stars using the fact that their isotropically smeared images have zero ellipticity ($\chi^{*,iso}=0$),
\begin{equation}
q_{\alpha}=(P^{*,sm})^{-1}_{\alpha \beta}\chi^{*,obs}_{\beta}
\end{equation}
Once $q$ has been determined, we can use Eq.~(\ref{eq:chiisoobs}) to compute the \textit{isotropic} from the \textit{observed} ellipticity. For a detailed calculation we refer again to \citep{Bartelmann01}. In the derivation, all the terms containing moments of $q$ higher than the second have been neglected as well as quadratic and higher-order terms in $q_{ij}$. If one wants to extend this calculation to higher orders in $q$, derivatives of the observed surface brightness $I^{obs}$ would appear in the calculation because the assumed equality to $I^{iso}$ (hypothetical surface brightness for vanishing $q$) does not hold any more. The fact that derivatives of the observed profile need to be considered renders it practically impossible to incorporate higher-order corrections for $q$, since $I^{obs}$ is a noisy quantity. This means that within the KSB framework it is not possible to correct properly for highly elliptical PSFs.

If the determination of $q$ is wrong, so is the estimate of $\chi^{iso}$ (the ellipticity of $I^{iso}$), and the error will propagate to the final shear measurement in an almost unpredictable way \citep{Erben01,KK99}. This could happen if the anisotropy of the PSF is too large for a linear treatment, or if the PSF cannot be decomposed into an isotropic and an anisotropic part.

\subsection{Tests}
\label{sec:psf_tests}
We perform the same tests as in the previous Section, but with an additional convolution with a Moffat-shaped PSF,
\begin{equation}
P(r)=(1+\alpha r^2)^{-\beta},
\end{equation}
where
\begin{equation}
\alpha=\frac{2^{1/\beta -1}}{(\mathrm{FWHM}/2)^2}
\end{equation}
controls the size of the PSF and $\beta$ regulates its steepness. In order to ensure vanishing flux at large radii, the PSF is truncated at $5$~FWHM, and the appropriate value at that position is subtracted from $P(r)$.

\begin{figure}
 	\centerline{
      \psfig{figure=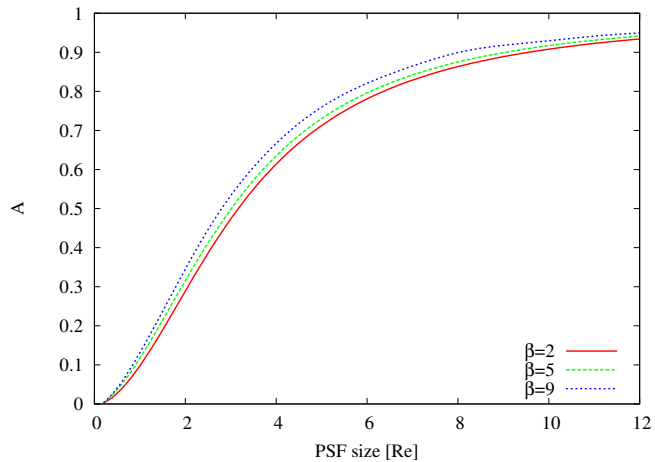,width=9cm,angle=270}}
\caption{Dependence of $A$ as defined in Eq.~(\ref{eq:A}) on the size of the PSF for a fixed value of the observed ellipticity.}\label{fig:APSF}
\end{figure}

We begin studying the case of a flat weight function, $W(x)=1$, for which we derive the behaviour of the four KSB variants in Eq.~(\ref{eq:gchiA}). The key quantity for describing a spherical PSF is given by $A$ as defined in Eq.~(\ref{eq:A}), which is a function mainly of the PSF width and mildly of its steepness for a given galaxy (see Fig.~\ref{fig:APSF}). We investigate the performance of the four methods as a function of the shear for a fixed PSF width. We choose $\mathrm{FWHM}=0.5\;R_e$ and $\beta=2$ to mimic a space-based observation, and $\mathrm{FWHM}=5\;R_e$ and $\beta=5$ to mimic a ground-based observation. The results are shown in Fig.~\ref{fig:PSF_nowin}. In the first case, KSB3 gives the best result, while KSBtr is the best approximation in the second case, as expected from Eq.~(\ref{eq:bias}). 
 
\begin{figure}
 	\centerline{
       \psfig{figure=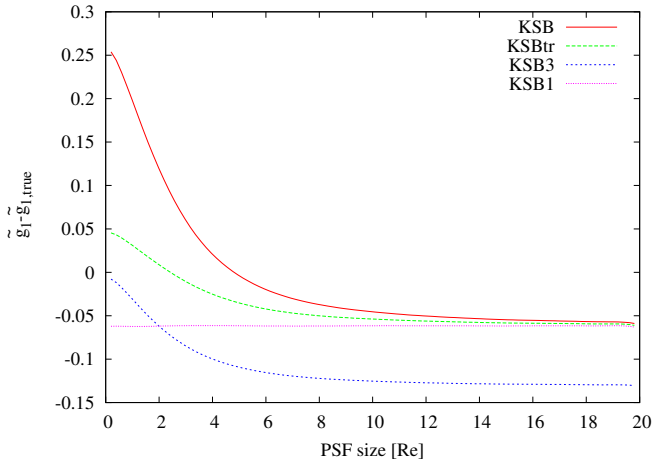,width=9cm,angle=270}}
\caption{Shear estimate $\tilde{g}_1$ as a function of the PSF size for a S\'ersic-type galaxy image as provided by KSB (red line), KSBtr (green line), KSB3 (blue line), and KSB1 (magenta line) for a fixed value of the preconvolved ellipticity corresponding to $\tilde{g}=(0.4,0.1)$).}\label{fig:PSF_size}
\end{figure}

We next investigate the response of the four methods to the size of the PSF for a given $\tilde{g}$ (Fig.~\ref{fig:PSF_size}). We choose $\tilde{g}_{1}=0.4$ and $\tilde{g}_{2}=0.1$. As expected from Eq.~(\ref{eq:bias}), KSB, KSBtr and KSB1 have the same limit for large PSF ($A\rightarrow 1$), while the bias for KSB3 is the largest in the limit of a very wide PSF. As noted before, the PSF correction in all KSB variants introduces a negative bias which partly compensates (or even overcompensates) the overestimate by KSB and KSBtr from the lens mapping. Since KSB3 is essentially unbiased for unconvolved ellipticities, any PSF correction necessarily lowers the shear estimate.

Finally, we introduce the weight function into the moment measurement and study the response of the four methods in this situation. The result is shown in Fig.~\ref{fig:PSF_win} for a space-based (left panel) and a ground-based observation (right panel). For narrow PSFs, the methods react on weighting as in the previous Section, where the PSF was neglected (see Fig.~\ref{fig:weightSize}), while the response is milder for a wider PSF. For a narrow PSF, KSBtr and KSB3 are essentially unbiased, and KSBtr remains fairly unbiased when the PSF width increases. 
From the comparison between Figs.~\ref{fig:PSF_nowin} and \ref{fig:PSF_win} 
we can infer the effect of weighting on the shear estimates. The biases of most methods are lowered because the ellipticity of the convolved source is lower, hence a circular weight function does not significantly affect the ellipticity measurement. However, in particular KSB1 shows concerning dependence on both the presence of a weighting function and the width of the PSF: Even though KSB1 seems fairly unbiased in the right panel of Fig.~\ref{fig:PSF_win}, other values of the width of the weight function would lead to less optimal results.

We are aware that our tests are of somewhat approximate nature in the sense that the characteristics of the simulated images only coarsely resemble that of realistic survey data. The real-life performance of all KSB variants will depend on peculiar properties of the surveys to be analyzed, such as the shape of the PSF, the depth of the observation, etc. However, two findings from our result can be considered robust: KSB3 shows the least amount of bias and the weakest dependence on the width of the weighting function, as long as the PSF remains narrow with respect to the galaxy size. KSBtr has a more pronounced dependence on the weighting function, but reacts only weakly on changes of the PSF width.

\section{Conclusions}

We have assessed the assumptions underlying the KSB method for measuring gravitational shear from the images of ensembles of lensed galaxies. KSB has the great advantage of being model-independent since it expresses the lensing-induced shape change by a combination of moments of the surface-brightness distribution. However, several assumptions underlying the derivation of the method and its practical implementations turn out to be violated more or less severely in realistic situations. We can summarise our results as follows:

\begin{figure*}
 \begin{minipage}{170mm}
 	\centerline{
       \psfig{figure=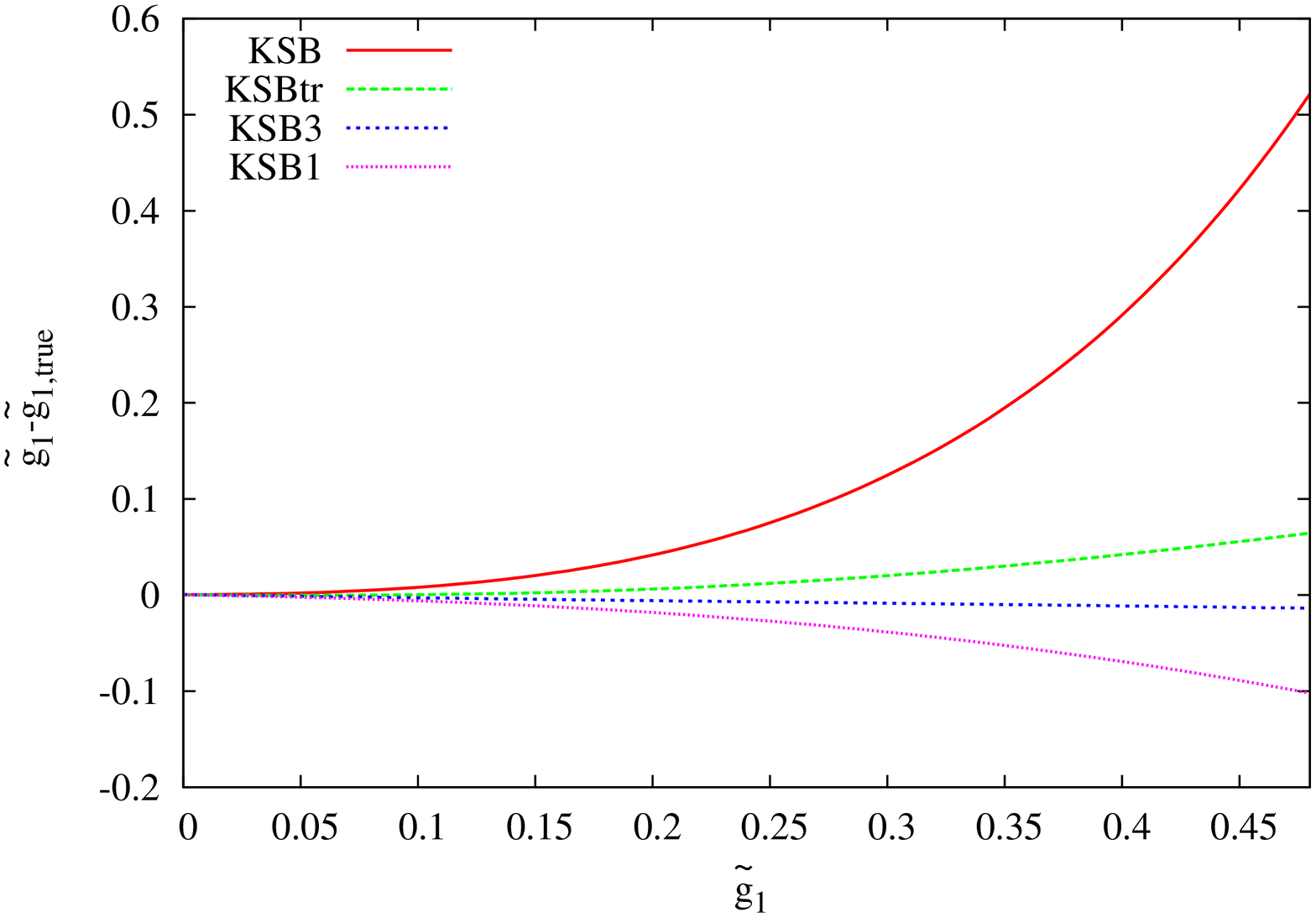,width=9cm,angle=270}
       \psfig{figure=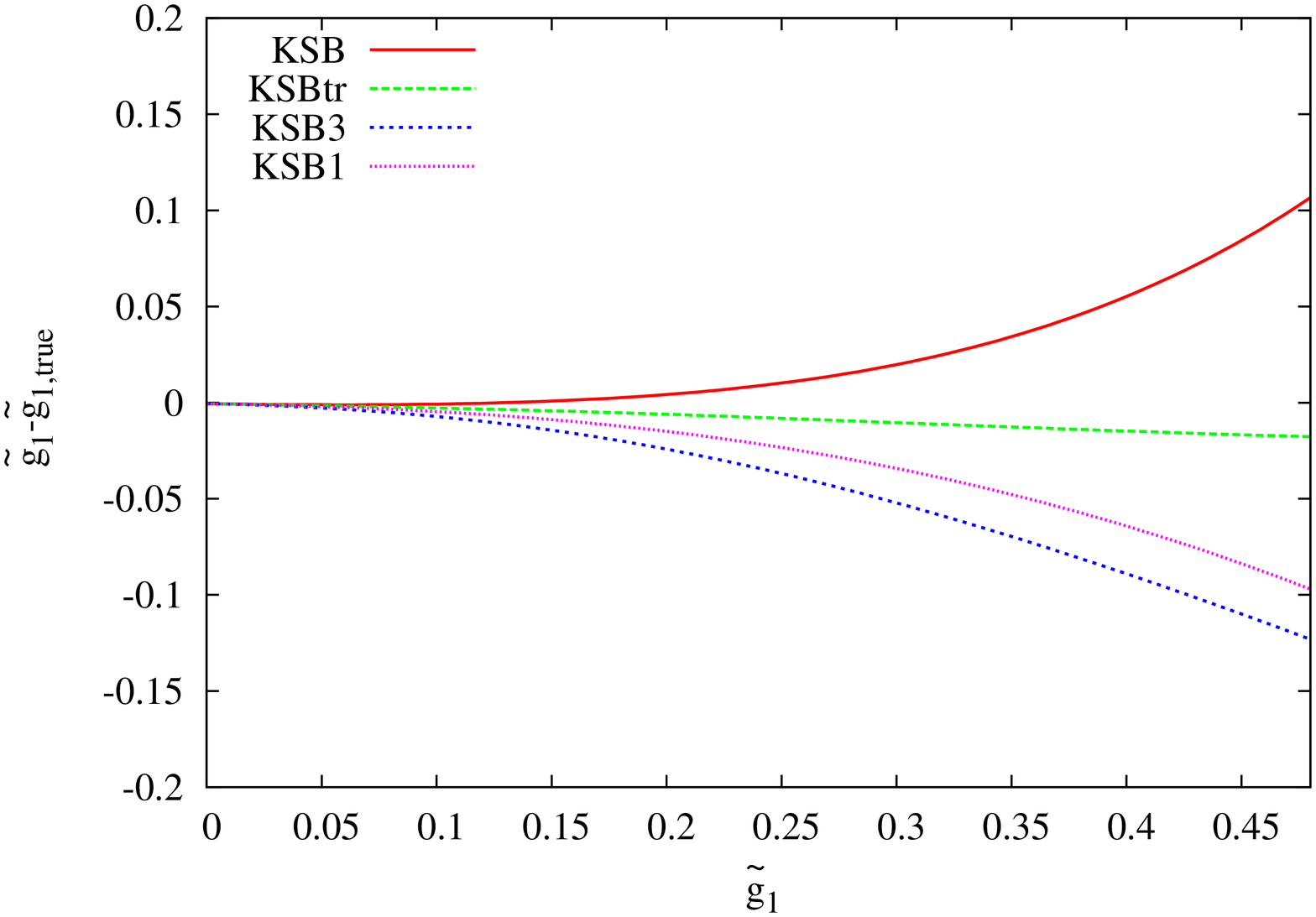,width=9cm,angle=270}}
\caption{Shear estimate $\tilde{g}_1$ as a function of the applied shear for noise-free S\'ersic-type galaxy images as provided by KSB (red line), KSBtr (green line), KSB3 (blue line), and KSB1 (magenta line). In the \textit{left panel} we choose a PSF with $\mathrm{FWHM}=0.5R_e$ and $\beta=2$ to mimic a space-based observation, while in the \textit{right panel} we choose $\mathrm{FWHM}=5R_e$ and $\beta=5$ to mimic a ground-based observation. No weight function has been used to compute moments.}\label{fig:PSF_nowin}
\end{minipage}
\end{figure*}

\begin{figure*}
 \begin{minipage}{170mm}
 	\centerline{
       \psfig{figure=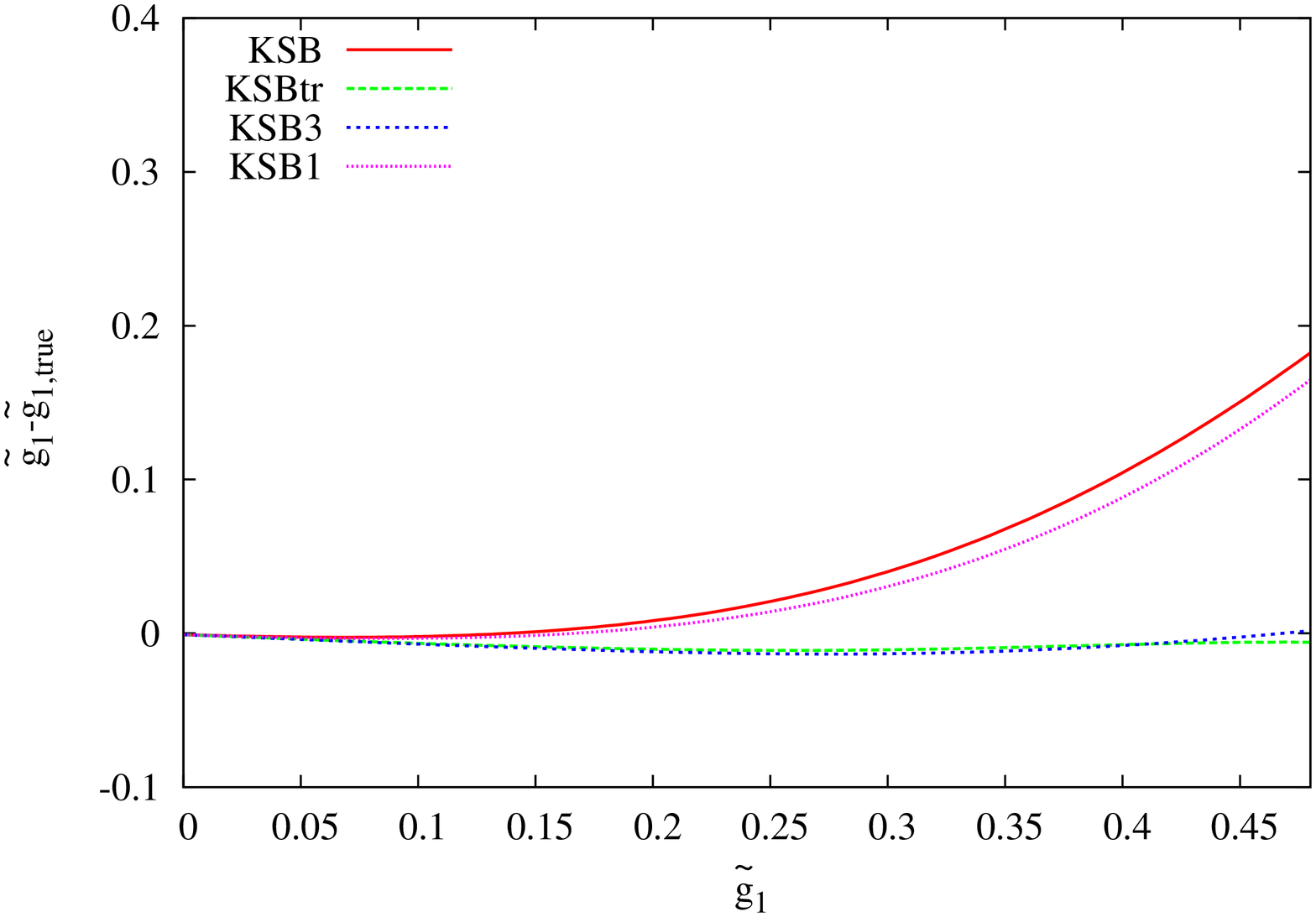,width=9cm,angle=270}
       \psfig{figure=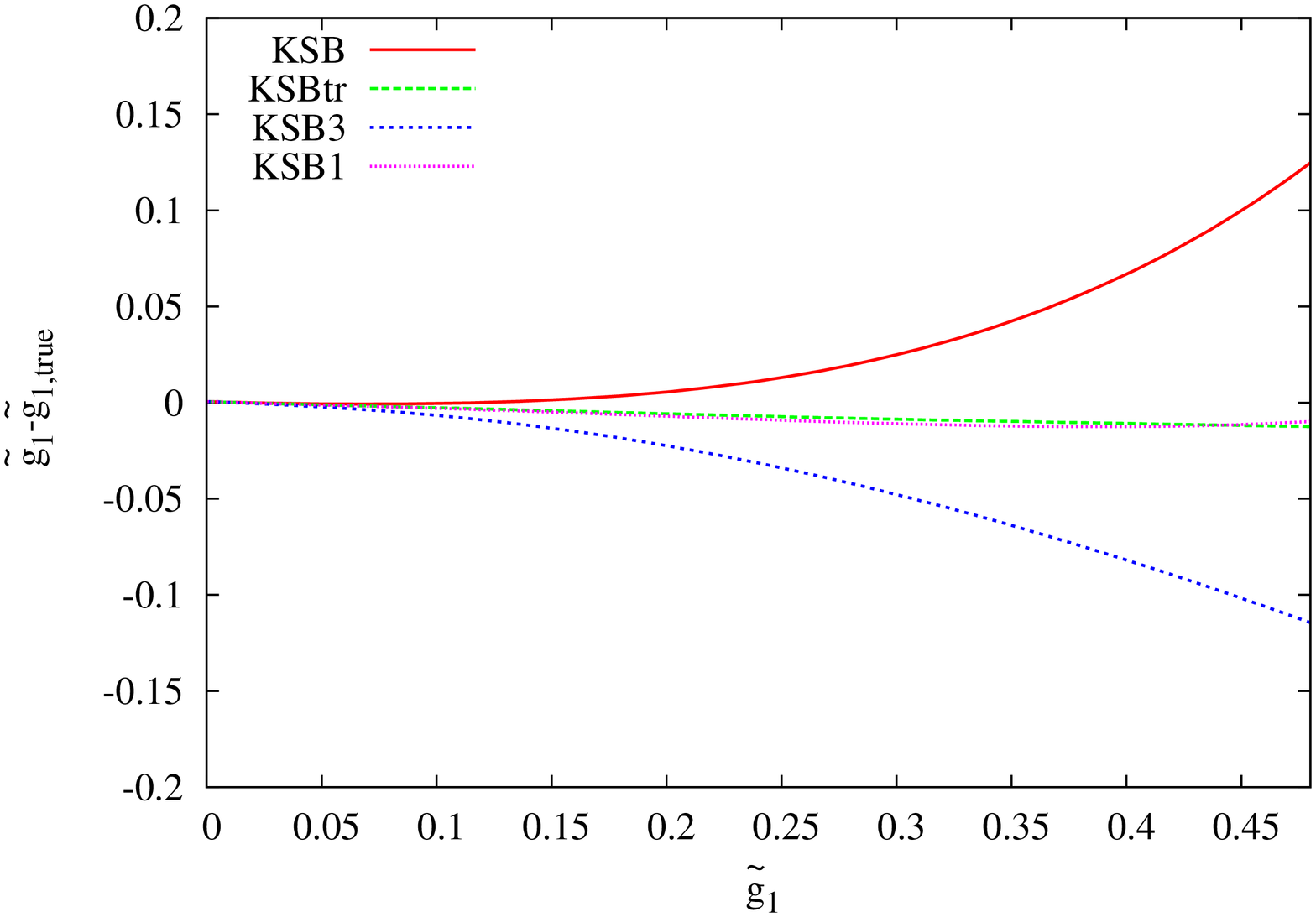,width=9cm,angle=270}}
\caption{As in Fig. \ref{fig:PSF_nowin} but employing a weighting function in the moments computation. The width of the weighting function was set to the apparent size of the objects, $\sigma = \sqrt{Tr(Q)}$}\label{fig:PSF_win}
\end{minipage}
\end{figure*}

\begin{enumerate}

\item KSB defines a shear estimate for each individual galaxy, defined as the shear that would describe the observed ellipticity if the object was perfectly circular prior to lensing. In other words, it is assumed that the instrinsic ellipticity of the individual object vanishes. The true shear is then computed averaging these shear estimates within a region where $g$ is assumed to be constant. This is in general not equivalent to averaging the ellipticities of each individual object and then computing the true shear: averaging observed galaxy ellipticities and measuring the shear do not commute because not the individual intrinsic ellipticities can be assumed to vanish, but only their average. We show that the difference between the two approaches is a function of the variance of the intrinsic ellipticity distribution. The error introduced this way depends on the variants of KSB used, the size of the PSF, and the width of the weight function. It is normally in the percent range.

\item The definition of the KSB shear estimate relies on the assumption that the shear is small. However, this is only true after averaging. For a single object, the reduced-shear estimate $g$ is of the same order as the ellipticity $\chi$. This leads to a relation between $g$ and $\chi$ which is correct only to first order. This situation can be improved considering only linear terms in $\chi g$ in the derivation of $P^{sh}$ (KSB1), or considering consistently terms up to third order in $\chi g$ (KSB3). We also show that the approximation of $P^{sh}$ by half of its trace (KSBtr), although not mathematically justified, yields a better $g$-$\chi$ relation compared to KSB.

\item KSB, KSB1 and KSBtr in absence of PSF convolution tend to overestimate the shear, while KSB3 gives an almost perfect result.

\item KSB and KSB1 depend strongly on the width of the weight function used in the moment measurements, while KSBtr is more robust and KSB3 is almost independent of it.

\item KSB does not perform any PSF deconvolution, but gives only an approximate correction for the effects of the PSF. We show that this correction would be equivalent to a proper deconvolution from a circular PSF only in the case of a circular source, otherwise the improper PSF correction lowers the shear estimate.

\item The overestimate due to the wrong relation between $g$ and $\chi$ and the underestimate due to the inappropriate PSF correction tend to compensate each other. For a narrow PSF (space-based observation), KSB3 is the variant with the least bias, while KSBtr is the best method for wider PSFs (ground-based observation).

\item The choice of the width $\sigma$ of the weight function could be utilized to reduce the measurement bias. In principle, $\sigma$ can be tuned according to the size of the PSF and to the galaxy ellipicities such that the shear estimate ends up to be almost unbiased. However, practically this is only feasible for the galaxy ensemble as a whole, whereas choosing $\sigma$ such that shear estimates are unbiased for each individual galaxy is of similar difficulty as estimating the shear.

\item KSB can correct only small anisotropies in the PSF ($q \ll 1$). It is not possible to extend the formalism to allow  more precise corrections since that would imply the calculation of derivatives of the observed surface brightness, which is not feasible since $I^{obs}$ is a noisy quantity. An improper correction of the PSF anisotropy introduces a bias which propagates to the final measurement of the shear in an almost unpredictable way.

\end{enumerate}

\appendix
\section{}

In this Appendix, we list expressions for the tensors defined in the paper in terms of moments of the light distribution: 

\begin{equation}
P_{11}=-\frac{2\chi_1L_{1}}{Tr(Q)}-2\chi^{2}_{1}+2\frac{B_{11}}{Tr(Q)}+2
\end{equation}
\begin{equation}
P_{12}=-2\frac{\chi_1 L_{2}}{Tr(Q)}-2\chi_{1}\chi_{2}+4\frac{B_{11}}{Tr(Q)}
\end{equation}
\begin{equation}
P_{22}=-2\frac{\chi_{2} L_{2}}{Tr(Q)}-2\chi^{2}_{2}+8\frac{Q^{\prime}_{1122}}{Tr(Q)}+2
\end{equation}
\begin{equation}
R_{111}=2\frac{K_{11}\chi_1}{Tr(Q)}+4\frac{B_{11}\chi_{1}}{Tr(Q)}-2\frac{D_{111}}{Tr(Q)}-4\frac{L_{1}}{Tr(Q)}
\end{equation}
\begin{equation}
R_{112}=2\frac{K_{12}\chi_1}{Tr(Q)}+4\frac{B_{12}\chi_1}{Tr(Q)}-2\frac{D_{112}}{Tr(Q)}
\end{equation}
\begin{equation}
R_{121}=2\frac{K_{12}\chi_1}{Tr(Q)}+4\frac{B_{12}\chi_1}{Tr(Q)}-2\frac{D_{112}}{Tr(Q)}-4\frac{L_{2}}{Tr(Q)}
\end{equation}
\begin{equation}
R_{211}=2\frac{K_{11}\chi_2}{Tr(Q)}+4\frac{B_{11}\chi_2}{Tr(Q)} -2\frac{D_{112}}{Tr(Q)}
\end{equation}
\begin{equation}
R_{122}=2\frac{K_{22}\chi_1}{Tr(Q)}+16\frac{Q^{\prime}_{1122}\chi_1}{Tr(Q)}-2\frac{D_{122}}{Tr(Q)}
\end{equation}
\begin{equation}
R_{221}=2\frac{K_{12}\chi_2}{Tr(Q)}+4\frac{B_{12}\chi_2}{Tr(Q)}-2\frac{D_{122}}{Tr(Q)}
\end{equation}
\begin{equation}
R_{212}=2\frac{K_{12}\chi_2}{Tr(Q)}+4\frac{B_{12}\chi_2}{Tr(Q)}-2\frac{D_{122}}{Tr(Q)}-4\frac{L_{1}}{Tr(Q)}
\end{equation}
\begin{equation}
R_{222}=2\frac{K_{22}\chi_2}{tr(Q)}+16\frac{Q^{\prime}_{1122}\chi_2}{tr(Q)}-16\frac{Q^{\prime \prime}_{111222}}{Tr(Q)}-4\frac{L_{2}}{Tr(Q)}
\end{equation}

\begin{equation}
L_{1}=Q^{\prime}_{1111}-Q^{\prime}_{2222}
\end{equation}
\begin{equation}
L_{2}=2(Q^{\prime}_{1112}+Q^{\prime}_{2221})
\end{equation}
\begin{equation}
B_{11}=Q^{\prime}_{1111}-2Q^{\prime}_{1122}+Q^{\prime}_{2222}
\end{equation}
\begin{equation}
B_{12}=B_{21}=2(Q^{\prime}_{1112}-Q^{\prime}_{1222})
\end{equation}
\begin{equation}
B_{22}=4Q^{\prime}_{1122}
\end{equation}

\begin{equation}
K_{11}=(Q^{\prime \prime}_{111111}-Q^{\prime \prime}_{111122}-Q^{\prime \prime}_{112222}+Q^{\prime \prime}_{222222})
\end{equation}
\begin{equation}
K_{12}=K_{21}=2(Q^{\prime \prime}_{111112}-Q^{\prime \prime}_{122222})
\end{equation}
\begin{equation}
K_{22}=4(Q^{\prime \prime}_{111122}+Q^{\prime \prime}_{112222})
\end{equation}
\begin{equation}
D_{111}=Q^{\prime \prime}_{111111}-3Q^{\prime \prime}_{111122}+3Q^{\prime \prime}_{111122}-Q^{\prime \prime}_{222222}
\end{equation}
\begin{eqnarray}
D_{112}&=&D_{121}=D_{211}= \nonumber \\
&=&2(Q^{\prime \prime}_{111112}-2Q^{\prime \prime}_{111222}+Q^{\prime \prime}_{222221})
\end{eqnarray}
\begin{equation}
D_{122}=D_{212}=D_{221}=4(Q^{\prime \prime}_{111122}-Q^{\prime \prime}_{112222})
\end{equation}
\begin{equation}
D_{222}=8Q^{\prime \prime}_{111222}
\end{equation}
\begin{eqnarray}
U_{1111}&=&Q^{\prime \prime \prime}_{11111111}-4Q^{\prime \prime \prime}_{11111122}+6Q^{\prime \prime \prime}_{11112222}\nonumber \\
&-&4Q^{\prime \prime \prime}_{11222222}+Q^{\prime \prime \prime}_{22222222}
\end{eqnarray}
\begin{eqnarray}
U_{2111}&=&2(Q^{\prime \prime \prime}_{11111112}-3Q^{\prime \prime \prime}_{11111222}\nonumber \\
&+&3Q^{\prime \prime \prime}_{11122222}-Q^{\prime \prime \prime}_{12222222})
\end{eqnarray}
\begin{equation}
U_{1211}=U_{1121}=U_{1112}=U_{2111}
\end{equation}
\begin{equation}
U_{2211}=4(Q^{\prime \prime \prime}_{11111122}-2Q^{\prime \prime \prime}_{11112222}+Q^{\prime \prime \prime}_{11222222})
\end{equation}
\begin{equation}
U_{2112}=U_{2121}=U_{1122}=U_{1221}=U_{1212}=U_{2211}
\end{equation}
\begin{equation}
U_{2221}=8(Q^{\prime \prime \prime}_{11111222}-Q^{\prime \prime \prime}_{11122222})
\end{equation}
\begin{equation}
U_{2122}=U_{2212}=U_{1222}=U_{2221}
\end{equation}
\begin{equation}
U_{2222}=16Q^{\prime \prime \prime}_{11112222}
\end{equation}
Note that the moments $Q^{\prime}_{ij..k}$ have to be computed using the first derivative of the weight function with respect to $\vec{\theta}^2$. Similarly, $Q^{\prime \prime}_{ij..k}$ must be computed using the second derivatives of the weight function.

\section*{Acknowledgements}

This work was supported by the EU-RTN ``DUEL'', the Heidelberg Graduate School of Fundamental Physics, the IMPRS for Astronomy \& Cosmic Physics at the University of Heidelberg and the Transregio-Sonderforschungsbereich TR~33 of the Deutsche Forschungsgemeinschaft. PM is supported by the DFG Priority Program 1177. We are grateful to Peter Schneider and Thomas Erben for useful comments and suggestions to improve the manuscript and we thank Julian Merten for carefully reading it.

\bibliographystyle{mn2e}

\end{document}